\documentclass[aps,prd,floatfix,superscriptaddress,preprintnumbers]{revtex4}

\usepackage{titlesec}
\usepackage{amssymb,amsmath,amsfonts,latexsym,graphicx,epsfig,fancybox}
\usepackage{empheq}
\usepackage{float}
\usepackage{xcolor}

\newcommand{\bea}{\begin{eqnarray}}
\newcommand{\ena}{\end{eqnarray}}
\newcommand{\nn}{\nonumber\\}
\newcommand{\be}{\begin{equation}}
\newcommand{\en}{\end{equation}}

\newcommand{\ed}{\end{document}}

\newcommand{\slp}{p\kern-5pt/}

\begin{document}

\hfill MITP/21-005

\title{Form-factor-independent test of lepton universality \\ 
in semileptonic heavy meson and baryon decays}

\author{Stefan~Groote}
\email{stefan.groote@ut.ee}
\affiliation{F\"u\"usika Instituut, Tartu \"Ulikool,
  W.~Ostwaldi 1, EE-50411 Tartu, Estonia}
\author {Mikhail~A.~Ivanov}
\email{ivanovm@theor.jinr.ru}
\affiliation{
Bogoliubov Laboratory of Theoretical Physics,
Joint Institute for Nuclear Research,
141980 Dubna, Russia}
\author{J\"urgen~G.~K\"{o}rner}
\email{jukoerne@uni-mainz.de}
\affiliation{PRISMA+ Cluster of Excellence, Institut f\"{u}r Physik, 
Johannes Gutenberg-Universit\"{a}t, 
D-55099 Mainz, Germany}
\author{Valery~E.~Lyubovitskij}
\email{valeri.lyubovitskij@uni-tuebingen.de}
\affiliation{Institut f\"ur Theoretische Physik,
Universit\"at T\"ubingen,
Kepler Center for Astro and Particle Physics,
Auf der Morgenstelle 14, D-72076 T\"ubingen, Germany}
\affiliation{Departamento de F\'\i sica y Centro Cient\'\i fico
Tecnol\'ogico de Valpara\'\i so-CCTVal, Universidad T\'ecnica
Federico Santa Mar\'\i a, Casilla 110-V, Valpara\'\i so, Chile}
\affiliation{Department of Physics, Tomsk State University,
634050 Tomsk, Russia}
\affiliation{Tomsk Polytechnic University, 634050 Tomsk, Russia}
\author{Pietro~Santorelli}
\email{Pietro.Santorelli@na.infn.it}
\affiliation{
  Dipartimento di Fisica ``E.~Pancini'', Universit\`a di Napoli Federico II,
  Complesso Universitario di Monte S.~Angelo,
  Via Cintia, Edificio 6, 80126 Napoli, Italy}
\affiliation{
  Istituto Nazionale di Fisica Nucleare, 
  Sezione di Napoli, 80126 Napoli, Italy}
\author{Chien-Thang Tran}
\email{thangtc@hcmute.edu.vn}
\affiliation{
  Dipartimento di Fisica ``E.~Pancini'', Universit\`a di Napoli Federico II,
  Complesso Universitario di Monte S.~Angelo,
  Via Cintia, Edificio 6, 80126 Napoli, Italy}
\affiliation{
  Istituto Nazionale di Fisica Nucleare,
  Sezione di Napoli, 80126 Napoli, Italy}
\affiliation{
    Department of Physics, Faculty of Applied Sciences,
  HCMC University of Technology and Education,
  Vo Van Ngan 1, Thu Duc, Ho Chi Minh City, Vietnam
    }

\date{\today}

\begin{abstract}

In the semileptonic decays of heavy mesons and baryons the lepton-mass
dependence factors out in the quadratic $\cos^2\theta$ coefficient of the
differential $\cos\theta$ distribution. We call the corresponding normalized
coefficient the convexity parameter. This observation opens the path to a test
of lepton universality in semileptonic heavy meson and baryon decays that is
independent of form-factor effects. By projecting out the quadratic rate
coefficient, dividing out the lepton-mass-dependent factor and restricting the
phase space integration to the $\tau$ lepton phase space, one can define
optimized partial rates which, in the Standard Model, are the same for all
three $(e,\mu,\tau)$ modes in a given semileptonic decay process. We discuss
how the identity is spoiled by New Physics effects. We discuss semileptonic
heavy meson decays such as $\bar{B}^0 \to D^{(\ast)+} \ell^- \bar\nu_\ell$ and
$B_c^- \to J/\psi (\eta_c)\ell^- \bar\nu_\ell$, and semileptonic heavy baryon
decays such as $\Lambda_b \to \Lambda_c \ell^- \bar\nu_\ell$ for each
$\ell=e,\mu,\tau$.

\end{abstract}

\maketitle

\section{Introduction}
\label{sec:intro}

Recently there has been an extraordinary amount of experimental and
theoretical activity on the analysis of semileptonic heavy meson and baryon
decays. The semileptonic decays  $B\to D^{(\ast)} + \ell\bar\nu_\ell$
($D^{(\ast)} = D\,\text{or}\, D^\ast$, $\ell=e,\mu,\tau$) are the best studied
processes.  Starting with the {\it BABAR}
papers~\cite{Lees:2012xj,Lees:2013uzd},
this upsurge of  activity has been fuelled by possible observations of the
violation of lepton flavor universality which, if true, would signal possible
New Physics (NP) contributions in these decays.
The  decays $B\to D^{(\ast)} + \tau\bar\nu_\tau$ have  been  also studied  by
the Belle \cite{Bozek:2010xy,Huschle:2015rga,Sato:2016svk,Hirose:2016wfn}
and LHCb  \cite{Aaij:2015yra} experiments.
The present situation concerning the so-called flavor anomalies is
summarized in
Refs.~\cite{Amhis:2019ckw,Bernlochner:2021vlv,Barbieri:2021wrc,Cheung:2020sbq}.

The present tests of lepton flavor universality suffer from their dependence
on the assumed form of the $q^2$ behavior of the transition form factors. In
the Standard Model (SM) the three semileptonic $(\ell=e,\mu,\tau)$ modes
of a given decay are governed by the same set of form factors.
However, due to the kinematical constraint $m^2_\ell \le q^2 \le (m_1-m_2)^2$
the form factors are probed in different regions of $q^2$. Furthermore,
the helicity flip factor
$\delta_\ell=m^2_\ell/2q^2$ multiplying the helicity flip contributions
provides an additional weight factor depending on $q^2$ and the lepton mass
which differ for the three modes. All in all, the tests of lepton universality
based on rate measurements alone suffer from a complex interplay of the above
two effects which is difficult to control. Ultimately, such tests require the
exact knowledge of the $q^2$ behavior of the various transition form factors
which is difficult to obtain with certainty (see, e.g.,
Ref.~\cite{Cohen:2019zev}). Instead, one would prefer tests of lepton
universality which are independent of form factor effects such as we are
proposing in this paper.

It turns out that the above two obstacles to a clean test of lepton
universality can be overcome by (i) restricting the analysis to the phase space
of the $\tau$ mode, and (ii) choosing angular observables for which the helicity
flip contributions can be factored out. Fortunately, such an observable is
provided by the coefficient of the $\cos^2\theta$ contribution in the
differential $\cos\theta$ distribution.

The restriction to a reduced phase space will lead to a loss in rate for the
$\ell=\mu,e$ modes which will hopefully be compensated by the 40-fold increase
in luminosity provided by the SuperKEKb accelerator at the Belle II detector.
For example, the loss in rate through the phase space reduction
$(\Gamma_{\rm tot}-\Gamma_{\rm red})/\Gamma_{\rm tot}$ is given by
${\cal O} (50) \%$ and ${\cal O} (30) \%$ for the decays
$\bar{B}^0 \to D^+ \ell^- \bar\nu_\ell$ and
$\bar{B}^0 \to D^{\ast+} \ell^- \bar\nu_\ell$ ($\ell=e,\mu $), respectively.
Much more demanding in terms of experimental accuracy is the fact that the
proposed test requires an angular analysis which is not mandatory
in those tests using the rate alone.
%

The proposed test of lepton universality will lead to the SM 
equality of certain optimized (``optd'') rates $\Gamma^{\rm optd}_{U-2L}$ 
in the three $(e,\mu,\tau)$ modes, i.e. one has
\be
\label{eq:LU}
\Gamma^{\rm optd}_{U-2L}(e)=\Gamma^{\rm optd}_{U-2L}(\mu)
  =\Gamma^{\rm optd}_{U-2L}(\tau).
\en
While the actual values of the optimized rates in Eq.~(\ref{eq:LU}) are
form-factor dependent,
the unit ratio of any of the two optimized rates in
Eq.~(\ref{eq:LU}) or, equivalently, the ratio of the corresponding branching
fractions is form-factor independent, i.e., one has
\be
\label{Rtaumu}
R^{\rm optd}(\ell,\ell')
  =\frac{\Gamma^{\rm optd}_{U-2L}(\ell)}{\Gamma^{\rm optd}_{U-2L}(\ell')}
  =\frac{B^{\rm optd}_{U-2L}(\ell)}{B^{\rm optd}_{U-2L}(\ell')}= 1.
  \en
  In this way one can test $\mu/e$, $\tau/\mu$ and $\tau/e$ lepton universality
  regardless of form-factor effects.  
  NP contributions designed to strengthen the $\tau$
  rate will clearly lead to a
violation of the equalities~(\ref{eq:LU}) or the unit ratio of optimized
rates~(\ref{Rtaumu}). The size of the NP violations can be used to constrain
the parameter space of the NP contributions in a model-dependent way.


\section{ Generic differential {\boldmath $\cos\theta$} distribution}
\label{sec:cos}


We discuss three kinds of semileptonic heavy hadron decays involving the
$b \to c$ current transition, namely the decays $P(0^-)\to P'(0^-)\ell\bar\nu$,
$P(0^-)\to V(1^-)\ell\bar\nu$, and $B(1/2^+)\to B'(1/2^+)\ell\bar\nu$. We
expand the generic differential $(q^2,\cos\theta)$ distribution for these
decays in terms of their helicity structure functions~\cite{Korner:1989ve,%
Korner:1989qb,Bialas:1992ny,Ivanov:2015tru,Kadeer:2005aq,Gutsche:2015mxa,%
Groote:2019rmj,DiSalvo:2018ngq}
\be
\label{q2cosdistr}
\frac{d^2\Gamma}{dq^2 \, d\cos\theta}
  =\,\frac{2}{2S_1 +1}\,\,\frac{3}{8}\frac{\Gamma_0 |\vec q\,|q^2 v^2}{m_1^7}
\Big(A_0 +A_1\cos\theta +A_2 \cos^2\theta\Big),
\en
where $S_1$ is the spin of the initial hadron, 
\bea 
\Gamma_0=\frac{G_F^2 |V_{cb}|^2 m_1^5}{192 \pi^3}
\ena 
is the fundamental rate occurring in the weak three-body decay 
transitions of particle with mass $m_1$ and governed by the weak 
coupling $G_F |V_{cb}|$. 
The momentum transfer is denoted by $q=p_1-p_2$, and
$|\vec q\,|=|\vec p_2\,| =\sqrt{Q_+Q_-}/2m_1$ is the
momentum of the daughter particle in the rest system of the parent particle 
with $Q_\pm=(m_1\pm m_2)^2-q^2$.
The polar angle of the charged lepton in the $(l,\nu_l)$ c.m.\
system relative to the momentum direction of the $W_{\rm{off-shell}}$ is denoted
by $\theta$.

The coefficients $A_0$, $A_1$, and $A_2$ are given by
\bea
A_0 &=&{\cal H}_U + 2{\cal H}_L +2\delta_\ell \, ({\cal H}_U + 2 {\cal H}_S), \\
    A_1 &=&-2\Big({\cal H}_P + 4\delta_\ell{\cal H}_{SL}\Big), \\
\label{eq:A2}
    A_2 &=& v \, ({\cal H}_U - 2{\cal H}_L)\,.
\ena
In (\ref{eq:A2}) we have introduced the velocity-type parameter
$v=1-m_\ell^2/q^2$ which, when expressed in terms of the helicity
flip factor $\delta_\ell=m_\ell^2/2q^2$, reads $v=1-2\delta_\ell$. The helicity
structure functions
${\cal H}_{X}\,\,(X=U,L,\ldots)$ are bilinear combinations of the helicity
amplitudes which will be specified later on. Note that the coefficient
$A_2$ factors into the $q^2$ and lepton-mass-dependent factor $v=1-m_\ell^2/q^2$,
and the $q^2$ dependent combination
${\cal H}_U(q^2) - 2{\cal H}_L(q^2)$. We mention
that instead of expanding the $(q^2,\cos\theta)$ distribution in terms of
helicity structure functions as in (\ref{q2cosdistr}) one can
also expand the decay distribution in terms of invariant
structure functions~\cite{Fischer:2001gp,Penalva:2019rgt,Penalva:2020xup,%
Penalva:2020ftd}.

The cosine of the polar angle
$\theta$ can be related to the energy $E_\ell$ of the lepton measured in the
rest system of the parent particle. The relation
reads (see, e.g.,~\cite{Korner:1989ve,Kadeer:2005aq})
\begin{equation}
\label{costheta}
\cos\theta =\frac{2E_\ell-q_0(1+2\delta_\ell)}{|\vec q\,|\,v}
\end{equation}
with $-1 \le \cos\theta \le 1$. The energy of the off-shell $W$ boson in
the rest system of the parent particle is given by  
\begin{equation}
q_0=(m_1^2-m_2^2+q^2)/(2m_1).
\end{equation}

For our purposes, it is more convenient to rewrite the $\cos\theta$
distribution in terms of the Legendre polynomials. One of the advantages of
the Legendre representation is that
one can project out the coefficient $A_2$ in a straightforward way.
One has
\begin{equation}
\label{eq:distr3}
\frac{d^2\Gamma}{dq^2 d\cos\theta}
=\frac{1}{2S_1+1}\frac{\Gamma_0|\vec q\,|q^2v^2}{m_1^7}
  \bigg\{{\cal H}_{\rm tot}(q^2,m^2_\ell) P_0(\cos\theta)
  + {\cal H}_1(q^2,m^2_\ell) P_1(\cos\theta)
  +  v{\cal H}_2(q^2) P_2(\cos\theta)\bigg\}.
\end{equation}
The coefficient functions ${\cal H}_{\rm tot}$, ${\cal H}_1$, and
${\cal H}_2$ are given by 
\bea\label{H_ampl}
{\cal H}_{\rm tot}(q^2,m_\ell^2)&=&
(1+\delta_\ell)( {\cal H}_U + {\cal H}_L) +3\delta_\ell\,{\cal H}_S,
  \nn
{\cal H}_1(q^2,m_\ell^2)&=&
-\frac32\,\Big({\cal H}_P +4\,\delta_\ell{\cal H}_{SL}\Big),
  \nn
{\cal H}_2(q^2)&=& \frac{1}{2} \, ( {\cal H}_U-2 {\cal H}_L ) 
= \frac{1}{2} \, {\cal H}_{U-2L}  \,. 
\ena
For the convenience of the reader we list some properties of the Legendre
polynomials,
\be
P_0(\cos\theta)=1, \qquad
P_1(\cos\theta)=\cos\theta, \qquad
P_2(\cos\theta)= \frac 12\,(3\cos^2\theta-1).
\en
The Legendre polynomials satisfy the orthonormality relation
\be
\label{eq:ortho}
\int_{-1}^{+1} dx P_m(x)P_n(x)=\frac{2}{2n+1}\delta_{mn}\,.
\en 
It is now straightforward to extract the observables ${\cal H}_{\rm tot}$,
${\cal H}_1$, and ${\cal H}_2$ from Eq.~(\ref{eq:distr3}) by folding the
angular distribution with the relevant Legendre polynomial. For instance, the
differential decay rate is obtained by folding in $P_0(\cos\theta)$
as follows:
\be
\frac{d\Gamma}{dq^2} = \int\limits_{-1}^1\! d\cos\theta
  \frac{d^2\Gamma}{dq^2 d\cos\theta} P_0(\cos\theta)
=\frac{2}{2S_1 +1}\frac{\Gamma_0 |\vec q\,|q^2 v^2}{m_1^7}
{\cal H}_{\rm tot}(q^2,m^2_\ell).
\label{eq:distr1}
\en
The partial differential rate $d\Gamma_{U-2L}/dq^2$ can be projected out
by folding in $ P_2(\cos\theta)$ according to
\be
\frac{d\Gamma_{U-2L}}{dq^2} =
10\int\limits_{-1}^1\! d\cos\theta
  \frac{d^2\Gamma}{dq^2 d\cos\theta} P_2(\cos\theta)
  =\frac{2}{2S_1 +1}\frac{\Gamma_0 |\vec q\,|q^2 v^3}{m_1^7}
  {\cal H}_{U-2L}(q^2),
\label{eq:H2}
\en
where the helicity structure function
${\cal H}_{U-2L}(q^2)$ defined in Eq.~(\ref{H_ampl}) is a function of $q^2$ only 
[see also Refs.~\cite{Penalva:2019rgt,Penalva:2020xup,Penalva:2020ftd}].
The overall factor $10$ in Eq.~(\ref{eq:H2}) has been chosen such to have
the same normalization of Eq.~(\ref{eq:distr1}) and Eq.~(\ref{eq:H2}).

In Refs.~\cite{Ivanov:2015tru,Gutsche:2015mxa} we have defined a convexity parameter
$C_F(q^2,\ell)$ as a measure of the curvature of the $\cos\theta$ distribution
by taking the second derivative of the $\cos\theta$ distribution. The relation
of the convexity parameter to the ratio of the two differential
rates (\ref{eq:distr1}) and (\ref{eq:H2}) is given by
\be
C_F(q^2,\ell)=\frac 34 \,\,\frac{d\Gamma_{U-2L}(q^2,\ell)/dq^2}{d\Gamma(q^2,\ell)/dq^2}\,.
\en
Also, we introduce the average values of the convexity
parameter $\langle C_F^{\ell} \rangle$ where the average is taken in the interval
$m_\tau^2 \le q^2 \le (m_1-m_2)^2$ for both $\mu$ and $\tau$ 
modes: 
\bea
\langle C_F^{\ell} \rangle
= \frac 34 \,\, 
\frac{\int_{m_\tau^2}^{(m_1-m_2)^2} \, dq^2 \, d\Gamma_{U-2L}(q^2,\ell)/dq^2}
{\int_{m_\tau^2}^{(m_1-m_2)^2} \, dq^2 \, d\Gamma(q^2,\ell)/dq^2}\,, 
\qquad \ell = \mu,\tau .
\ena 
An interesting method to compare the theoretical prediction
    for the angular observables like the convexity parameter  
    with experimental data was proposed in Ref.~\cite{Kim:2018hlp}.
    It is based on counting the number of events in certain regions of
    the Dalitz plot.


\section{\label{sec:optd}Optimized observables}       

\begin{figure}[H]
\centering  
\epsfig{figure=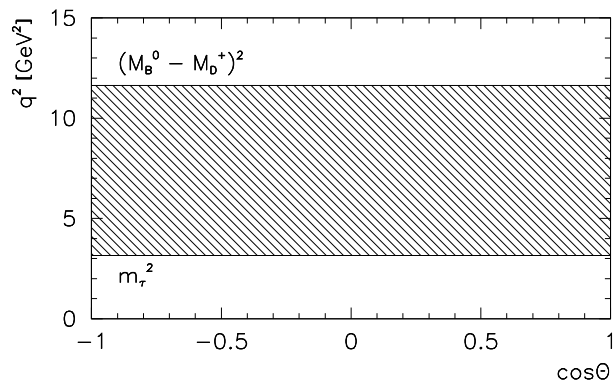, scale=0.50}
\caption{\label{figpspace}$(q^2,\cos\theta)$ phase space for
$\bar B^0 \to D^{+} +\ell^- +\bar{\nu}_\ell$. The hatched
region shows the $\ell=\tau$ phase space.}
\end{figure}

The possible breaking of lepton flavor universality is usually studied by
analyzing the ratios of rates or, equivalently, the ratio of branching ratios
for the tau and muon modes. As discussed in the introduction one can remove
the lepton-mass effects by introducing two improvements. First,
we propose to analyze observables in the common phase space region
$m^2_\tau\le q^2\le (m_1-m_2)^2$ as has been suggested before in
Refs.~\cite{Freytsis:2015qca,Bernlochner:2016bci, Isidori:2020eyd}.
As an example, in Fig.~\ref{figpspace} we show the $(q^2,\cos\theta)$ phase
space for the decay $\bar B^0 \to D^{+} +\ell^- +\bar{\nu}_\ell$ where the
hatched area shows the common phase space region
$m^2_\tau\le q^2\le (m_1-m_2)^2$. Second, we reweigh suitable observables in
which the lepton-mass dependence factors out by dropping the overall
lepton-mass-dependent factor. As Eqs.~(\ref{q2cosdistr})
and~(\ref{eq:distr3}) show,
such an observable is available through the coefficient of the quadratic
$\cos^2\theta$ term in the angular decay distribution proportional to the
helicity structure function $v{\cal H}_{U-2L}$. 
       
Based on Eq.~(\ref{eq:H2}) we define an optimized differential partial rate
by dividing out the factor $v^3$. One has 
\be
\label{eq:U-2L_Goptd}
\frac{d\Gamma^{\rm optd}_{U-2L}(q^2,\ell)}{dq^2}
=v^{-3}\frac{d\Gamma_{U-2L}(q^2,\ell)}{dq^2}
= \frac{2\Gamma_0}{2S_1+1}\frac{|\vec q\,|q^2}{m_1^7}\,{\cal H}_{U-2L}(q^2),
\en
which by construction does not depend on the lepton mass. In terms of the
ratios of branching fractions
\bea 
B^{\rm optd}_{U-2L}(q^2,\ell) = \tau \, 
\frac{d\Gamma^{\rm optd}_{U-2L}(q^2,\ell)}{dq^2} \,,
\ena
where $\tau$ is the lifetime of the respective hadron, 
Eq.~(\ref{eq:U-2L_Goptd}) leads to
\be
\label{eq:U-2L_RBoptd}
R^{\rm optd}_{U-2L}(q^2;\ell,\ell')=\frac{B^{\rm optd}_{U-2L}(q^2,\ell)}
{B^{\rm optd}_{U-2L}(q^2,\ell')}=1.
\en
Eq.~(\ref{eq:U-2L_RBoptd}) can be used to test lepton universality on the
differential $q^2$ level by analyzing the ratios of the optimized branching
fractions $R^{\rm optd}_{U-2L}(q^2;\tau,\mu)
=R^{\rm optd}_{U-2L}(q^2;\tau,e)=R^{\rm optd}_{U-2L}(q^2;\mu,e)$ in the reduced
phase space region $m_\tau^2 \le q^2 \le q^2_{\rm max}$. In practice one would
lump the light lepton modes together and concentrate on the ratio of branching
fractions $R^{\rm optd}_{U-2L}(q^2;\tau,(\mu+e))=1/2$.

After $q^2$ integration over the reduced phase space region one has
\be
\Gamma^{\rm optd}_{U-2L}(\ell)=\int_{m^2_\tau}^{(m_1-m_2)^2}\,dq^2
\frac{d\Gamma^{\rm optd}_{U-2L}(q^2,\ell)}{dq^2}\,.
\en
The proposed test of lepton universality will lead to the equality of the
optimized partial rates
$\Gamma^{\rm optd}_{U-2L}(\ell)$ in the three $(e,\mu,\tau)$ modes,
\be
\Gamma^{\rm optd}_{U-2L}(e)=\Gamma^{\rm optd}_{U-2L}(\mu)
  =\Gamma^{\rm optd}_{U-2L}(\tau)
\en
or, equivalently, to the equality of the three corresponding optimized
branching ratios
 \be
B^{\rm optd}_{U-2L}(e)=B^{\rm optd}_{U-2L}(\mu)
  =B^{\rm optd}_{U-2L}(\tau).
\en 
The equality of the three optimized rates or optimized branching ratios is
independent of form-factor effects, while the actual value of the optimized
rates or optimized branching ratios is form-factor dependent and is thus model
dependent. However, the ratio of the $(e,\mu,\tau)$ branching fractions are
predicted to be equal to one, independently of form-factor effects,
i.e., one has 
\be
\label{Rratios}
R^{\rm optd}_{U-2L}(\ell,\ell')= \frac{B^{\rm optd}_{U-2L}(\ell)}
  {B^{\rm optd}_{U-2L}(\ell')}=1.
\en

Since the $(q^2,\cos\theta)$ phase space is rectangular the $q^2$ and
$\cos\theta$ integrations can be interchanged. One can therefore first
integrate over $q^2$ and then do the $U-2L$ projection rather than first
projecting out ${\cal H}_{U-2L}$ and then doing the $q^2$ integration. This may
be of advantage in the experimental analysis.

Note that our definition of the optimized rates or branching ratios differs
from the one used in Ref.~\cite{Isidori:2020eyd}. In order to differentiate
between the two definitions we denote our optimized rates by the label
``${\rm optd}$'' instead of the label ``${\rm opt}$'' used in
Ref.~\cite{Isidori:2020eyd}. The authors of Ref.~\cite{Isidori:2020eyd} define
an optimized rate ratio
\be
\label{isidori}
R^{\rm opt}=\frac{\int_{m_\tau^2}^{(m_1-m_2)^2}d\Gamma^{\rm opt}(\tau)/dq^2}
{\int_{m_\tau^2}^{(m_1-m_2)^2}(1-2\delta_\tau)^2(1+\delta_\tau)
d\Gamma^{\rm opt}(\mu)/dq^2}\,>\,1\,.
\en
The numerator exceeds the denominator because of
the addition of a definitely positive scalar contribution in the numerator.

The idea behind the definition (\ref{isidori}) is to define an R-measure
$R^{\rm opt}$ which minimizes the propagation of form factor errors to the
optimized R-measure $R^{\rm opt}$. This goal is, in fact, achieved by the
R-measure $R^{\rm opt}$ (\ref{isidori}) as shown in
Ref.~\cite{Isidori:2020eyd}.


\section{\label{threeclasses} Three classes of semileptonic decays}


We now discuss three classes of prominent $b \to c$ induced semileptonic
decays in turn. We begin with the decay $P(0^-)\to P'(0^-)\ell\bar\nu_\ell$. 

\subsection{\boldmath{$P(0^-)\to P'(0^-)\ell\bar\nu_\ell$} decay} 
The decays $\bar{B}^0 \to D^{+} \ell^- \bar\nu_\ell$ and
$B_c^+ \to \eta_c\, \ell^+\nu_\ell$ belong to this class of decays.
The two form factors describing the $B\to D$ transition are defined by
(see, e.g., Refs.~\cite{Korner:1989ve,Ivanov:2015tru})
\be
\label{covariant1}
\langle P_2|J_\mu^{V}|P_1\rangle= F_+(q^2)(p_1 +p_2)_\mu + F_-(q^2)q_\mu\,.
\en
The corresponding helicity amplitudes $H_{\lambda_W}$ read
\be
H_0 = \frac{2m_1 |\vec q\,|}{\sqrt{q^2}}\,F_+(q^2), \qquad
H_\pm=0, \qquad H_t = \frac{1}{\sqrt{q^2}}( m_+ m_- F_+(q^2) + q^2 F_-(q^2) ),
\en
where $m_\pm=m_1\pm m_2$.

The longitudinal and scalar helicity structure functions are given in terms
of the bilinear combinations 
\be
 {\cal H}_L = |H_0|^2, \qquad  {\cal H}_U= |H_+|^2 + |H_-|^2= 0, \qquad
 {\cal H}_S=|H_t|^2.
\en

Note that the longitudinal structure function ${\cal H}_L$ is proportional to
$|\vec q\,|^2$. Since the unpolarized transverse structure function
${\cal H}_U$ is zero, one has ${\cal H}_{U-2L} \sim |\vec q\,|^2$.

\subsection{\boldmath{$P(0^-)\to V(1^-)\ell\bar\nu_\ell$} decay} 
Interesting decays in this class are
$\bar{B}^0 \to D^{\ast\,+} \ell^- \bar \nu_\ell$ and
$B_c^- \to J/\Psi\, \ell^- \bar \nu_\ell$.
We define invariant form factors according to the expansion 
(see, e.g., Refs.~\cite{Korner:1989ve,Ivanov:2015tru})
\be
\label{c0variant2}
\langle V_2|J_\mu^{V-A}|P_1\rangle=\frac{\epsilon^{\dagger\,\alpha}}{m_+}
\biggl(-g_{\mu \alpha} Pq A_0(q^2)+P_\mu P_\alpha A_+(q^2)
+q_\mu P_\alpha A_-(q^2) +i \varepsilon_{\mu\alpha P q}V(q^2)\,\biggr) \,. 
\en

One has to specify the helicity amplitudes $H_{\lambda_W\lambda_V}$ by the two
helicities $\lambda_W$ and $\lambda_V$ of the off-shell $W$ boson and the
daughter vector meson. The helicity amplitudes are given by
\bea
H_{t0} &=& \frac{m_1\,m_-\,|{\vec q}\,|}{m_2\sqrt{q^2}} \, 
\biggl(-A_0+A_+ + \frac{q^2}{m_+ m_-} \, A_-\biggr) \,,
\nn[1.2ex]
H_{\pm1\pm1} &=& m_- \biggl( - \, A_0 \pm \frac{2\,m_1}{m_+m_-} \,|{\vec q}\,|\, V \biggr) \,,
\nn[1.2ex]
H_{00} &=&  \frac{m_-}{2 \,m_2\,\sqrt{q^2}} 
\biggl(- (m_+m_- - q^2)\, A_0 + 
\frac{4\,m_1^2}{m_+m_-} \,|{\vec q}\,|^2\, A_+\biggr) \,.
\label{eq:hel_vv}
\ena
The helicity structure functions read  
\be 
{\cal H}_U=|H_{+1+1}|^2+|H_{-1-1}|^2, \qquad {\cal H}_L=|H_{00}|^2,
\qquad {\cal H}_S=|H_{t0}|^2.
\en
Note that ${\cal H}_S,\,{\cal H}_{U-2L}\sim |\vec q\,|^2$.
This scaling is obvious for ${\cal H}_S$. In the case of ${\cal H}_{U-2L}$,
it requires a little algebra based on the use of the identity:
\bea 
|{\vec q}\,|^2 = \frac{(m_+m_- - q^2)^2}{4 m_1^2} 
- \frac{m_2^2}{m_1^2} q^2 \,. 
\ena 

\subsection{\boldmath{$B(\frac12^+)\to B'(\frac12^+)\ell\bar\nu_\ell$} decay}

One defines the invariant form factors by writing (see, e.g.,  
Refs.~\cite{Gutsche:2015mxa,Groote:2019rmj})
\bea 
\label{covariant3}
\langle B_2|J_\mu^{V/A}|B_1\rangle =
  \ \ \bar{u}_p(p_2)\bigg[F_1^{V/A}(q^2)\gamma_\mu
  -i\frac{F_2^{V/A}(q^2)}{m_1}\sigma_{\mu \nu}q^\nu
  +\frac{F_3^{V/A}(q^2)}{m_1}q_\mu\bigg] \, 
(I/\gamma_5) \, u_n(p_1) \,.
\ena
The corresponding helicity amplitudes $H^{V/A}_{\lambda_2\lambda_W}$ read
\bea
H_{\frac12\,t}^{V/A}&=&\frac{\sqrt{Q_\pm}}{\sqrt{q^2}}
\bigg( m_\mp F_1^{V/A}(q^2) \pm \frac{q^2}{m_1} F_3^{V/A}(q^2) \bigg) \,,
\nn
H_{\frac12\,0}^{V/A}&=&\frac{\sqrt{Q_\mp}}{\sqrt{q^2}}
\bigg( m_\pm F_1^{V/A}(q^2) \pm \frac{q^2}{m_1} F_2^{V/A}(q^2) \bigg) \,,
\nn
H_{\frac12\,1}^{V/A}&=&\sqrt{2Q_\mp}
\bigg(F_1^{V/A}(q^2) \pm \frac{ m_\pm}{m_1} F_2^{V/A}(q^2) \bigg) \,. 
\ena
From parity or from an explicit calculation one has
$ H^{V}_{-\lambda_2-\lambda_W} = + H^{V}_{\lambda_2\lambda_W} $ and
$ H^{A}_{-\lambda_2-\lambda_W} = - H^{A}_{\lambda_2\lambda_W} $. The relevant
helicity structure functions read
\begin{equation}
{\cal H}_U = 2\,\Big( |H^V_{+\frac12\, +1}|^2 + |H^A_{+\frac12\, +1}|^2 \Big)\,, \qquad
{\cal H}_L = 2\,\Big( |H^V_{+\frac12\, 0}|^2 + |H^A_{+\frac12\, 0}|^2   \Big)\,, \qquad
{\cal H}_S = 2\,\Big( |H^V_{+\frac12\, t}|^2 + |H^A_{+\frac12\, t}|^2   \Big)\,.
\end{equation}
With a little algebra one finds ${\cal H}_{U-2L}\sim |\vec q\,|^2$.

In all three classes of decays one finds that the helicity structure function 
combination ${\cal H}_{U-2L}={\cal H}_U-2{\cal H}_L$ is proportional to
$|\vec q\,|^2$. This leads to a depletion of the partial rate
$d\Gamma^{\rm opt}_{U-2L}/dq^2$ close to the zero recoil $q^2=(m_1-m_2)^2$
where $|\vec q\,|=0$. In this paper we do not study the parity-odd
helicity structure functions ${\cal H}_P$ and ${\cal H}_{SL}$, which 
scale as ${\cal H}_P,{\cal H}_{SL} \sim|\vec q\,|$~\cite{Ivanov:2015tru,%
Kadeer:2005aq,Gutsche:2015mxa,Groote:2019rmj}. 


\section{Numerical results}


We are now in the position to discuss the numerical values for the optimized
observables introduced in our paper.
The key point here is the choice of the form factors characterizing
the $B\to D^{(\ast)}$ and $\Lambda_b\to\Lambda_c$ transitions.
In addition to various model calculations there are precise lattice QCD
determinations for these form factors.  The first lattice-QCD determination
of the form factors describing the semileptonic decays
$\Lambda_b\to \Lambda_c^{(\ast)} + \ell\bar\nu_\ell$ has been performed
in Refs.~\cite{Detmold:2015aaa,Datta:2017aue,Meinel:2021rbm}.
The Fermilab Lattice and MILC collaborations have presented the
computations of zero-recoil form factor for $B\to D^{(\ast)}+\ell\bar\nu_\ell$
decay in Ref.~\cite{Bailey:2014tva} and unquenched lattice-QCD calculation
of the hadronic form factors for the exclusive  decay $B\to D+\ell\bar\nu_\ell$
at  nonzero  recoil in Ref.~\cite{Lattice:2015rga}.
In Ref.~\cite{Na:2015kha} the HPQCD collaboration has presented  a lattice
QCD calculation of the $B\to D+\ell\bar\nu_\ell$ decay for the entire
physical $q^2$ range. The branching fraction ratio was found to be
$R(D)=0.300(8)$. The $B-D$ calculations in particular are precise, cover
various $q^2$ values and have been combined with experimental data for the
light lepton $q^2$ distribution to cover the full spectrum.
Something similar has been done with $B-D^\ast$ as well,
see  Refs.~\cite{Bigi:2016mdz,Bernlochner:2017jka,Gambino:2019sif,
       Bordone:2019vic}.  The lattice determinations of the
form factors were also employed to extract the value of $V_{cb}$.
 The numerical values for the optimized observables introduced in this paper
 are calculated by using the form factors obtained in the framework of
 the covariant confined quark  model (CCQM). The behavior of all CCQM form
 factors were found to be quite smooth
in the full kinematical range of the semileptonic transitions. In fact, they
are well represented by a two-parameter representation in terms of 
a double-pole parametrization 
\be
F(q^2) = \frac{F(0)}{1-a s +b s^2}, \qquad s=\frac{q^2}{m_1^2}.
\en
The values of the fitted parameters $a$,$b$ and $F(q^2=0)$ 
are listed in Eq.~(34) of Ref.~\cite{Ivanov:2015tru} for
the $B\to D^{(\ast)}$ transition, in Table~I of Ref.~\cite{Tran:2018kuv} for
the $B_c\to \eta_c$ and $B_c\to J/\psi$ transitions, and in Eq.~(59) of
Ref.~\cite{Gutsche:2015mxa} for the $\Lambda_b\to\Lambda_c$ transition. The
values of the lepton and hadron masses, their lifetimes as well as the value
of the CKM matrix element $V_{cb}$ are taken from the PDG~\cite{PDG2020}.

In Table~\ref{tab:conv-obs} we list the average values of the convexity
$\langle C_F^{\ell} \rangle$. 
For the two transitions $B \to D$ and $B_c \to \eta_c$ we get 
$\langle C_F^{\mu} \rangle = - 1.49 \simeq -3/2$ in the $\mu$-mode. 
The reason for such a common value in both transitions is that there are no
transverse contribution in the $P \to P'$ transitions and the muon mass is
strongly suppressed in comparison with the $\tau$ lepton mass
($m_\mu/m_\tau \ll 1)$. 
In the limit $m_\mu/m_\tau \equiv 0$ one gets 
$\langle C_F^{\mu} \rangle \equiv -3/2$. 
In case of the $\tau$-mode for the two $P \to P'$ transitions 
the average convexity parameter is quite small: $-0.26$ for the 
$B \to D$ transition and $-0.24$ for the $B_c \to \eta_c$ transition. 
Note that the entries in Table~\ref{tab:conv-obs} are form-factor dependent. 
In case of the $P \to V$ transitions one can see that 
the average convexity parameter is again suppressed for the $\tau$ modes. 
We also notice that $\langle C_F^\ell\rangle$ is more suppressed 
for the $P \to V$ transitions in comparison with the $P \to P'$ transitions. 
Finally, for the $\Lambda_b \to \Lambda_c$ transition we get 
the $\langle C_F^\ell\rangle$ parameters, which lie  between 
the ones for the $P \to V$ and $P \to P'$ transitions. 

\begin{table}[H]
\caption{$q^2$ averages of the convexity parameters
 $\langle C_F^{\mu} \rangle$ and $\langle C_F^{\tau} \rangle$ in the range
 $m_\tau^2 \le q^2 \le (m_1-m_2)^2$.
 \label{tab:conv-obs}
}
\vskip 3mm
\centering
\def\arraystretch{1.2}
\begin{tabular}{cccccc}
  \hline\hline
  Obs. \quad & \quad $B \to D$ \quad & $B_c \to \eta_c$\quad & $B\to D^\ast$
  & $B_c \to J/\psi$ & $\Lambda_b \to \Lambda_c$  \\
\hline
$\langle C_F^\mu\rangle$
  & $-1.49$ & $-1.49$ & $-0.27$ & $-0.22$ & $-0.44$ \\
$\langle C_F^\tau\rangle$
  & $-0.26$ & $-0.24$ & $-0.062$ & $-0.050$ & $-0.10$ \\
\hline\hline
\end{tabular}
\end{table}

In Fig.~\ref{fig:opt-hel} we show the behavior of
$d\Gamma^{\rm optd}_{U-2L}/dq^2$
and $d\Gamma_{U-2L}/dq^2=v^3\,d\Gamma^{\rm optd}_{U-2L}/dq^2$ ($\tau$-mode)
in the region $m^2_\tau\le q^2\le (m_1-m_2)^2$. In the case of $\ell,\mu$ modes
the two above rates coincide  with high accuracy.

The differential
rates are largest at threshold $q^2=m_\tau^2$ and go to zero at the
zero-recoil point $q^2=(m_1-m_2)^2$ with the characteristic $|\vec q\,|^3$ 
dependence.
The (form factor dependent) numerical values of the integrated observables
are given in Table~\ref{tab:opt-obs}. We also list their average values
for the range $4\, \text{GeV}^2 \le q^2\le (m_1-m_2)^2$ to highlight the fact
that the differential rates are largest in the region close to threshold
where, in the $\tau $-mode, the division
by $v^3$ is potentially problematic from the experimental
point of view.

Next we address the question of how to compare the numerical values calculated
in Table~\ref{tab:opt-obs} with the outcome of the corresponding experimental
measurements. We first assume that the number of the produced parent particles
is known which, in the case of produced $\bar{B}^0$'s, we will refer to as
$N(\bar{B}^0{\rm-tags})$. For example, in $e^+\,e^-$ annihilations on the
$\Upsilon(4S)$ resonance the bottom mesons are produced in pairs and the
identification of a $B^0$ on one side can be used as a tag for the $\bar{B}^0$
on the opposite side. In an experimental analysis one counts the number of
events of a given decay and relate this to the known number of produced
particles given by $N(\bar{B}^0{\rm-tags})$.

\begin{table}[ht]
  \caption{The optimized partial rate $\Gamma^{\rm optd}_{U-2L}$
    in  units of $10^{-14}\,\text{GeV}$.
    \label{tab:opt-obs}
  }
  \vskip 3mm
  \centering
\def\arraystretch{1.2}
\begin{tabular}{cccccc}
  \hline\hline
  $q^2_{\rm min}$&$B\to D$&$B_c\to\eta_c$&$B\to D^\ast$&$B_c\to J/\psi$
  & $\Lambda_b-\Lambda_c$ \\ \hline
$m^2_\tau$        & $-1.14$ & $-1.21$ & $-0.73$ & $-0.49$ & $-0.90$ \\
$4$ GeV$^2$       & $-0.89$ & $-0.93$ & $-0.54$ & $-0.36$ & $-0.71$ \\
\hline\hline
\end{tabular}
\end{table}

\begin{figure}[hb]
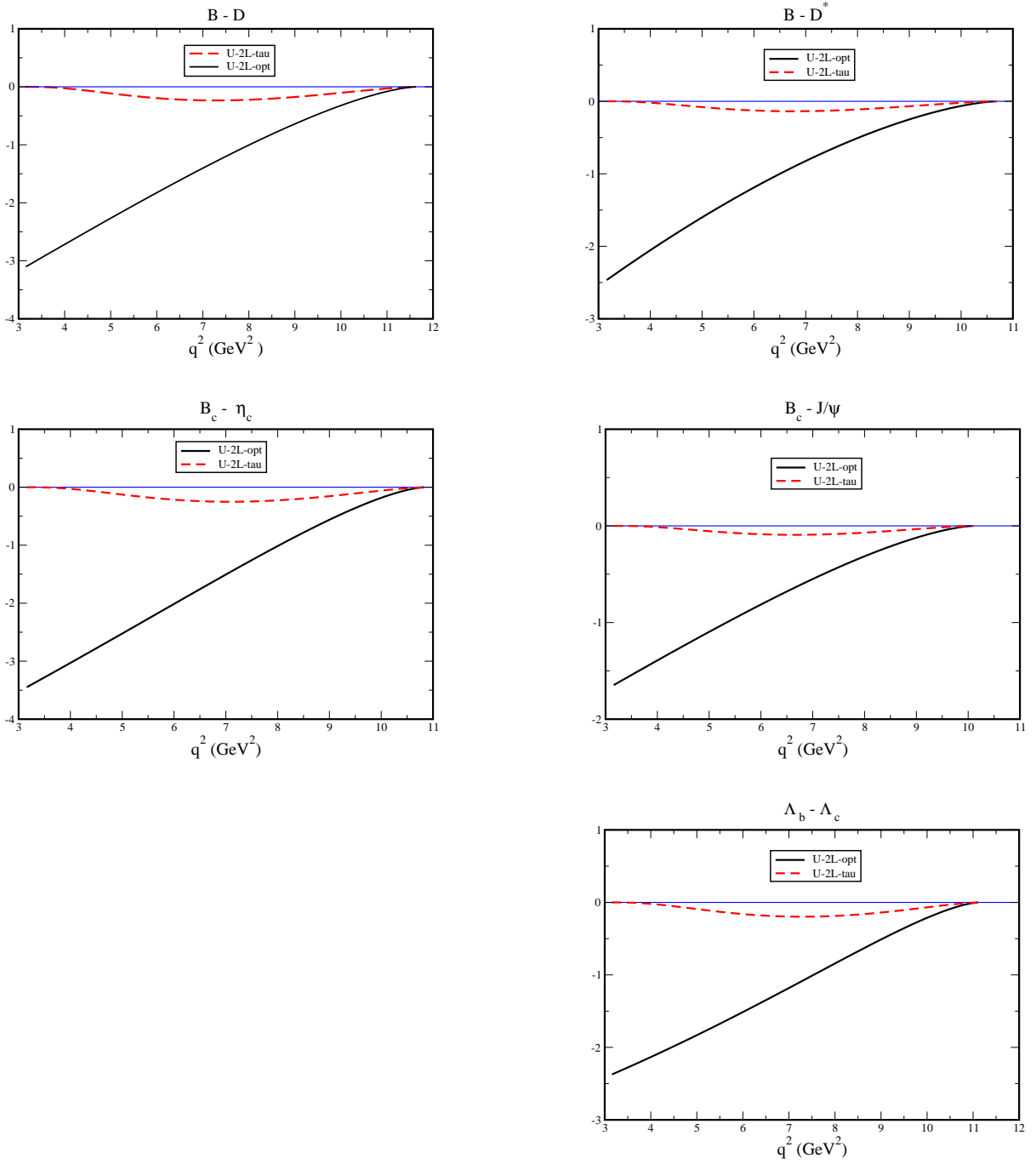

\vspace*{.5cm}
  \def\arraystretch{1.5}
  \begin{tabular}{lr}
   \includegraphics[scale=0.3]{fig2a_conv.eps} \hspace*{2cm} &
   \includegraphics[scale=0.3]{fig2b_conv.eps} \vspace*{0.5cm} \\
   \includegraphics[scale=0.3]{fig2c_conv.eps}  &
   \includegraphics[scale=0.3]{fig2d_conv.eps}\vspace*{0.5cm}\\
 & \includegraphics[scale=0.3]{fig2e_conv.eps} \\
\end{tabular}
\caption{$q^2$ dependence of the optimized partial rate
    $d\Gamma^{\rm optd}_{U-2L}/dq^2$ (solid curve) and
    $d\Gamma_{U-2L}/dq^2=v^3\,d\Gamma^{\rm optd}_{U-2L}/dq^2$
    ($\tau$-mode, dashed curve)
    in units of $10^{-15}$~GeV$^{-1}$.
\label{fig:opt-hel}    }
\end{figure}

One can then define an experimental branching fraction by writing
\be
\label{tag0}
B(\bar{B}^0 \to D^{+} \ell^- \bar{\nu}_{\ell})=
\frac{N(\bar{B}^0 \to D^{+} \ell^- \bar{\nu}_{\ell})}{N(\bar{B}^0{\rm-tags})}
\en
which can be compared to the theoretical branching fraction
\be
B(\bar{B}^0 \to D^{+} \ell^- \bar{\nu}_{\ell})=\tau(\bar{B}^0)
\Gamma_{\rm tot}(\bar{B}^0 \to D^{+} \ell^- \bar{\nu}_{\ell}).
\en
In the same way one can define an experimental optimized branching fraction
by writing
\be
\label{tag1}
B^{\rm optd}_{U-2L}(\bar{B}^0 \to D^{+} \ell^- \bar{\nu}_{\ell})=
  \frac{N^{\rm optd}_{U-2L}(\bar{B}^0 \to
  D^{+} \ell^- \bar{\nu}_{\ell})}{N(\bar{B}^0-{\rm tags})}
\en
which, again, can be compared to the corresponding theoretical branching
fraction
\be
B^{\rm optd}_{U-2L}(\bar{B}^0 \to D^{+} \ell^- \bar{\nu}_{\ell})=\tau(\bar{B}^0)
  \Gamma^{\rm optd}_{U-2L}(\bar{B}^0 \to D^{+} \ell^- \bar{\nu}_{\ell}).
\en

One then defines optimized rate ratios $R^{\rm optd}_{U-L}(\ell,\ell')$ by
\be
\label{optdrr}
R^{\rm optd}_{U-L}(\ell,\ell')= \frac{B^{\rm optd}_{U-2L}(\ell)}
{B^{\rm optd}_{U-2L}(\ell')}
=\frac{N^{\rm optd}_{U-2L}(\ell)}{N^{\rm optd}_{U-2L}(\ell')}
=\frac{\Gamma^{\rm optd}_{U-2L}(\ell)}{\Gamma^{\rm optd}_{U-2L}(\ell')}\,=\,1,
\en
which are predicted to be equal to one.

As the ratios~(\ref{optdrr}) show, tagging is not really required when
measuring the optimized rate ratio $R^{\rm optd}_{U-L}(\ell,\ell')$, since the
denominators $N(\bar{B}^0-{\rm tags})$ drop out when taking the
ratio~(\ref{optdrr}). This shows that the optimized rate ratio
$R^{\rm optd}_{U-L}(\ell,\ell')$ can be experimentally determined even for
untagged decays as in the $B_c^-$ and  $\Lambda_b$ decays.


\section{New Physics Contributions}


At present, $b-c$ transition puzzles motivate many studies of New Physics due
to the observed deviations from the Standard Model predictions.
  There is a number of theoretical attempts to resolve these puzzles.
  See, for example, Refs.~\cite{Ciezarek:2017yzh,Becirevic:2016hea} and
  other references therein.
Possible NP contributions to the semileptonic decays
$\bar{B}^0 \to D(D^\ast) \tau^- \bar{\nu}_{\tau}$
and $\bar{B}_c \to \eta_c(J/\psi) \tau^- \bar{\nu}_{\tau}$ have been studied
in our papers~\cite{Ivanov:2016qtw,Tran:2017udy,Tran:2018kuv}. The NP
transition form factors have been calculated in the full kinematic $q^2$
range employing again the CCQM.  The
modifications of the partial differential rates $d\Gamma_{U-2L}(\tau)/dq^2$
from the differential $(q^2,\cos\theta)$ distributions of the decays
$\bar B^0\to D\tau^-\bar\nu_\tau$ and $\bar B^0\to D^\ast\tau^-\bar\nu_\tau$
are presented in Eqs.~(14) and~(C1), respectively, in
Ref.~\cite{Ivanov:2016qtw}. One has
\be
\frac{d\Gamma_{U-2L}({\rm NP})}{dq^2} = \frac{2\Gamma_0}{2S_1+1}
\frac{|\vec q\,| q^2}{m_1^7}\, (1-2\delta_\tau)^3\, {\cal H}_{U-2L}({\rm NP})\,,
\en
where
\be
\label{eq:hel-NP}
{\cal H}_{U-2L}({\rm NP}) = \left\{\begin{array}{ll}
    -2|1+V_L+V_R|^2|H_0|^2 + 32|T_L|^2|H_T|^2 &
   \qquad (P-P')\text{-transition}\,,\\[1.5ex]
    (|1+V_L|^2+|V_R|^2)(|H_{++}|^2+|H_{--}|^2 - 2|H_{00}|^2) & \nn 
    -4 {\rm Re}V_R\, (H_{++}H_{--}- |H_{00}|^2) & \nn
    -16|T_L|^2(|H_T^+|^2+|H_T^-|^2 -2|H_T^0|^2) & \qquad (P-V)\text{-transition} \,.
   \end{array}\right.
\en
        If we recall the relations of helicities with the Lorentz form factors
        then one gets
        \bea
        \label{eq:ff-NP}
      {\cal H}_{U-2L+NP}^{P-P'} &=& \frac{4m_1^2|\vec q\,|^2}{q^2}
      \nn
      &\times& \Big\{ -2\, |1+V_L+V_R|^2\, F^2_+
      +32\, |T_L|^2\,\frac{q^2}{m_+^2}F^2_T\Big\} \nonumber 
      \\[2ex]
      |H_{++}|^2 + |H_{--}|^2 -2 |H_{00}|^2  &=&
      \frac{2m_1^2 |\vec q\,|^2}{m_2^2\,m_+^2\,q^2}
      \nn
      &\times& \Big\{-(Pq)^2\,A_0^2+2\Big[2m_2^2q^2V^2
        +Pq(Pq-q^2)A_0A_+\Big]       
      - 4m_1^2|\vec q\,|^2A_+^2\Big\} \,, 
       \nn[1.5ex]
      H_{++}H_{--} -|H_{00}|^2  &=&
      \frac{m_1^2 |\vec q\,|^2}{m_2^2\,m_+^2\,q^2}
      \nn
      &\times& \Big\{-(Pq)^2\,A_0^2-2\Big[2m_2^2q^2V^2
        -Pq(Pq-q^2)A_0A_+\Big]
      - 4m_1^2|\vec q\,|^2A_+^2\Big\} \,, 
     \nn[1.5ex]
     |H_T^+|^2 + |H_T^-|^2  - 2 |H_T^0|^2  &=&
      \frac{2m_1^2 |\vec q\,|^2}{m_2^2}
\Big\{ \frac{8m_2^2}{q^2}\,G_1^2- (G_1+G_2)^2
\nn
&+& \frac{2}{m_+^2}\Big[ (m_1^2+3m_2^2-q^2)\,G_1+(Pq-q^2)\,G_2\Big]G_0 
- \frac{4m_1^2|\vec q\,|^2}{m_+^4}\,G_0^2 \Big\}    \,.
\ena        

One can see that the differential rate $d\Gamma_{U-2L}$ vanishes
as $|\vec q\,|^3$ at zero recoil. 
Here, $V_{L/R}$ and $T_L$ are the complex Wilson coefficients governing the NP
contributions.
One has to note that the scalar operators contribute to
  the full four-fold angular distribution but they do not appear
  in the coefficient proportional to $\cos^2\theta$, i.e.,
  in the convexity parameter.
It is assumed that NP only affects leptons of the third
generation, i.e., the $\tau$ lepton mode. Note that the lepton-mass-dependent
factor $v$ also factors out in the NP contributions to the
$(U-2L)$ helicity structure function.

The parameters of the dipole approximation for the calculated NP form factors
are
listed in Eqs.~(10) and~(11) of Ref.~\cite{Ivanov:2016qtw} for $B-D$ and
$B-D^\ast$ transitions, and in Table~I of the Ref.~\cite{Tran:2018kuv} for
$B_c-\eta_c$ and $B_c-J/\psi$ transitions. The allowed regions for the NP
Wilson coefficients have been found by fitting the experimental data for the
ratios $R(D^{(\ast)})$ by switching on only one of the NP operators at a time.

\begin{figure}[H]
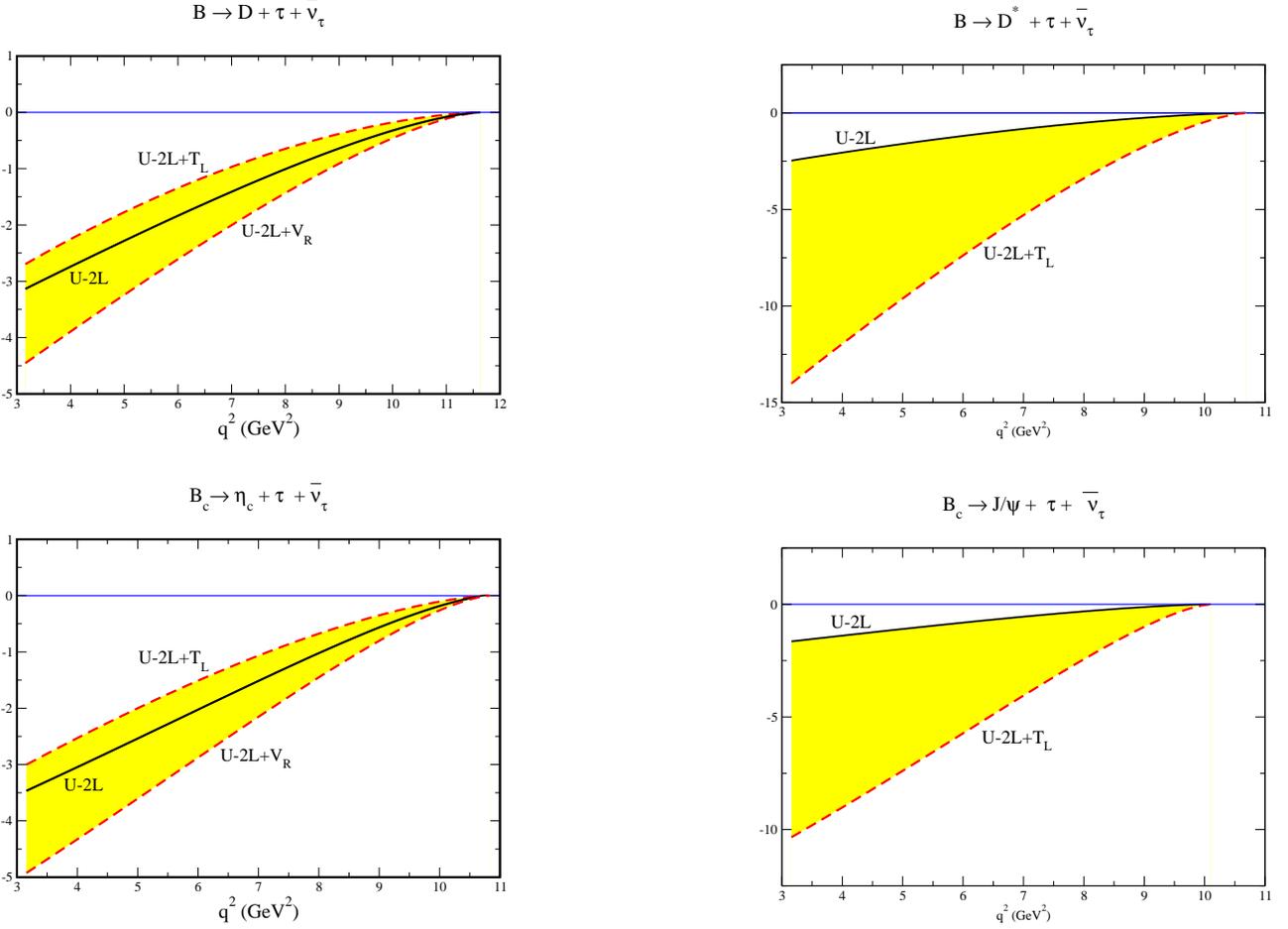

\vspace*{.5cm}
\def\arraystretch{1.25}
\begin{tabular}{lr}
\includegraphics[scale=0.3]{fig3a_conv.eps}    \hspace*{15mm}    &
\hspace*{15mm} 
\includegraphics[scale=0.3]{fig3b_conv.eps} 
\\[12pt]
\includegraphics[scale=0.3]{fig3c_conv.eps}        &
\includegraphics[scale=0.3]{fig3d_conv.eps}
\end{tabular}
\caption{\label{fig:SM+NP_opt-B-hel} $P\to P'\,(V)$ semileptonic transitions
taking into account NP effects for the $\tau$ mode. The $q^2$ dependence of
the optimized partial rates are shown in units of $10^{-15}$~GeV$^{-1}$. In the
figures we
make use of the short hand notation $U-2L = d\Gamma^{\rm optd}_{U-2L}/dq^2$.}
\end{figure}

In each allowed region at $2\sigma$ the best-fit value for each NP coupling
was found. The best-fit couplings read
\bea
\begin{aligned}
V_L &=-0.23-0.85i, \qquad & V_R &=0.03+0.60i,
\qquad & T_L &=0.38+0.06i.
\end{aligned}
\label{eq:bestfit}
\ena   
We define optimized rates for the NP contributions in the same way as has
been done for the SM in Eq.~(\ref{eq:U-2L_Goptd}). In
Fig.~\ref{fig:SM+NP_opt-B-hel} we plot the SM differential $q^2$ distributions
of the optimized rates $d\Gamma_{U-2L}^{\rm optd}/dq^2$ together with the
corresponding (SM+NP) distributions for the $\tau$-mode.
  In general, there are four curves for each mode. To avoid oversaturation
  of the figures, we display the upper and lower curves only
and the region between these two curves, colored in yellow.
The $P \to P'$ optimized differential rates are
enhanced by the NP $V_L$ and $V_R$ contributions, and reduced by the
NP tensor contribution $T_L$. For the $P \to V$ transitions the enhancement
due to the NP tensor contribution $T_L$ is quite pronounced over the whole
$q^2$ range.

The enormous size of the NP tensor contribution to the $P \to V$ transitions
also shows up in Table~\ref{tab:NP} where we list the integrated optimized
rates and the  $\tau/\mu$ ratio of optimized branching fractions 
\bea
R^{\rm optd}_{U-2L}(\tau,\mu) = \frac{\Gamma^{\rm optd}_{U-2L}({\rm SM+ NP})} 
{\Gamma^{\rm optd}_{U-2L}({\rm SM})} \,.
\ena 
The
deviations of the ratio of optimized branching fractions from the SM value of 1
is substantial and huge for the $P \to V$ transitions. One should be remindful
of the fact that the NP optimized $\tau$ rates and thereby the ratio of
branching fractions $B^{\rm optd}_{U-2L}(\tau,\mu)$ are form-factor dependent.

\begin{table}[ht]
  \caption{\label{tab:SM+NP_opt-obs}Optimized $(U-2L)$ rates in units of
    $10^{-14}$ GeV and rate ratios. NP effects are included in the
    $\tau$-mode only.
  \label{tab:NP}}
\vskip 3mm
\centering
\def\arraystretch{1.2}
\begin{tabular}{cccccc}
\hline\hline
Obs. & NP-coupling & $B\to D\ell\nu_\ell$      & $B_c\to \eta_c\ell\nu_\ell$
                   & $B\to D^\ast\ell\nu_\ell$ & $B_c\to J/\psi\ell\nu_\ell$ \\
\hline
$\Gamma^{\rm optd}_{U-2L}({\rm SM})$
&  & $-1.14$ & $-1.21$ & $-0.73$ & $-0.89$ \\
\hline
$\Gamma^{\rm optd}_{U-2L}({\rm SM+ NP})$
  & $V_L$ & $-1.50$ & $-1.59$ & $-0.96$ & $-0.64$ \\
  & $V_R$ & $-1.62$ & $-1.72$ & $-0.94$ & $-0.62$ \\
  & $T_L$ & $-0.85 $ & $-0.93$ &$-4.46$ & $-3.32$ \\
\hline
$R^{\rm optd}_{U-2L}(\tau,\mu)$
  & $V_L$ & $1.32$ & $1.31$ & $1.32$ & $0.72$ \\
  & $V_R$ & $1.42$ & $1.42$ & $1.29$ & $0.70$ \\
  & $T_L$ & $0.75$ & $0.77$ & $6.11$ & $3.73$ \\
\hline\hline
\end{tabular}
\end{table}


\section{Some concluding remarks}


As the authors of Ref.~\cite{Penalva:2019rgt} have emphasized, it is important
to also have a look at the $(q^2,E_\ell)$ distribution in semileptonic decays
when testing lepton universality. We briefly discuss the merits of using
the $(q^2,E_\ell)$ distribution for form-factor-independent tests of lepton
universality. One merit of $(q^2,E_\ell)$ distribution is obviously that
$\cos\theta$ is a derived quantity
whereas the lepton energy can be directly measured.

The $(q^2,\cos\theta)$ distribution~(\ref{q2cosdistr}) can be transformed to
the $(q^2,E_\ell)$ distribution by making use of the the
relation~(\ref{costheta}) between $\cos\theta$ and $E_\ell$. One obtains
\begin{equation}
\label{q2Eldistr}
\frac{d\Gamma}{dq^2dE_\ell}=\frac{1}{2S_1+1}\,\,
  \frac{3q^2}{|\vec q\,|^2}\frac{\Gamma_0}{m_1^7}
  \bigg(B_0(q^2,m_\ell)+B_1(q^2,m_\ell)\Big(\frac{E_\ell}{m_1}\Big)+B_2(q^2)
  \Big(\frac{E_\ell^2}{m_1^2}\Big)\bigg).
\end{equation}
where the coefficients $B_0(q^2,m_\ell),\,B_1(q^2,m_\ell)$,
and $B_2(q^2)$ are given by
\bea
\label{B0}
B_0(q^2,m_\ell)&=&\frac{1}{4}\Big(q_0^2(1 +2\delta_\ell)^2
  ({\cal H}_U-2{\cal H}_L)
  +v\,|\vec q\,|^2({\cal H}_U+2{\cal H}_L+2\delta_\ell
  ({\cal H}_U+2{\cal H}_S))\nn
&&+2q_0|\vec q\,|(1+2\delta_\ell)({\cal H}_P+4\delta_\ell{\cal H}_{SL})\Big), \\
\label{B1}
B_1(q^2,m_\ell)&=&-m_1\Big(q_0(1+2\delta_\ell)
  \left({\cal H}_U-2{\cal H}_L\right)
  +|\vec q\,|\left({\cal H}_P+4\delta_\ell{\cal H}_{SL}\right)\Big),\\
\label{B2}
  B_2(q^2)&=& \,m_1^2\,\Big({\cal H}_U-2{\cal H}_L\Big)\,.
\ena
The $(q^2,\cos\theta)$ distribution~(\ref{q2Eldistr}) can be seen to be well
defined in the limit $|\vec q\,| \to 0$ since
${\cal H}_P ,{\cal H}_{SL} \sim|\vec q\,|$ and 
${\cal H}_U-2{\cal H}_L\sim |\vec q\,|^2$ in all three classes of decays
as discussed in Sec.~\ref{threeclasses}. 

\begin{figure}[H]
  \epsfig{figure=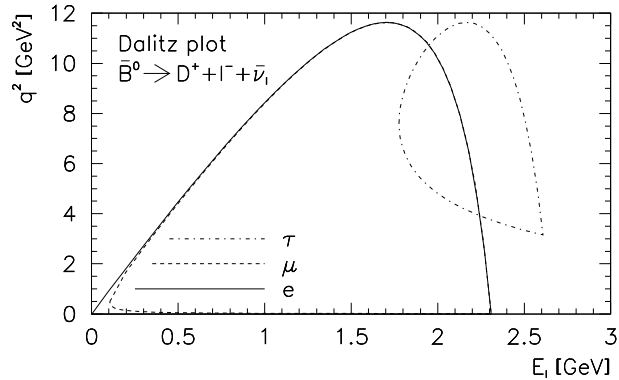, scale=0.50}
 \centering
 \caption{\label{dalitz2} $(q^2,E_\ell)$ phase space for
   $\bar B^0 \to D^{+} +\ell^- +\bar{\nu}_\ell$ for the three $(e,\mu,\tau)$
modes.}
\end{figure}

In Fig.~\ref{dalitz2} we show the $(q^2,E_\ell)$ phase space boundaries of the
three ($e,\mu,\tau $) modes of the semileptonic decay
$ \bar B^0 \to D^+ + \ell^-\,\bar \nu_\ell$. The phase space boundaries are
determined by the curves~\cite{Korner:1989qb,Kadeer:2005aq}
\begin{equation}
\label{q2-bound}
q^2_{\pm} = \frac{1}{a}\left(b \pm \sqrt{b^2-ac}\right),
\end{equation} 
where 
\begin{eqnarray}
a&=&m_1^2+m_\ell^2-2m_1E_\ell,\nonumber \\[5pt]
b&=&m_1E_\ell(m_1^2-m_2^2+m_\ell^2-2m_1E_\ell)+m_\ell^2m_2^2, \nonumber \\
c&=&m_\ell^2\Big((m_1^2-m_2^2)^2+m_\ell^2m_1^2-(m_1^2-m_2^2)2m_1E_\ell\Big).
\nonumber
\end{eqnarray}

From the relation~(\ref{costheta}) linking $\cos\theta$ and $E_\ell$ it is
not difficult to see that the coefficients of the $\cos^2\theta$ and
$E^2_\ell$ terms are simply related. In particular, as Eq.~(\ref{B2}) shows,
the coefficient $B_2(q^2)$ of the quadratic $E^2_\ell$ term is proportional to
${\cal H}_{U-2L}$ and, differing from the corresponding coefficient
${\cal H}_2$ of the $(q^2,\cos\theta)$ distribution, does not depend on the
lepton mass. A gratifying feature of the $(q^2,E_\ell)$ analysis is the fact
that the (model dependent) ratio $A_2(q^2)/A_0(q^2,m_\ell)$ is quite large
over the whole $q^2$ range~\cite{Penalva:2019rgt}.

Similar to Eq.~(\ref{eq:H2}), the second order coefficient
$B_2(q^2)=m_1^2{\cal H}_{U-2L}(q^2)$ can be projected from the
distribution~(\ref{q2Eldistr}) by folding the distribution with the second
order Legendre polynomial expressed in terms of the lepton energy, i.e.,
\be
P_2(\cos\theta(E_\ell))=\frac 32 \frac{1}{|\vec q\,|^2v^2}
  \Big(4E_\ell^2-4E_\ell q_0(1+2\delta_\ell)+q_0^2(1+2\delta_\ell)^2
  -\frac 13|\vec q\,|^2v^2\Big).
\en
The folding has to be done within the limits $(E_\ell^+,E_\ell^-)$
where (see, e.g., Refs.~\cite{Korner:1989qb,Kadeer:2005aq})
\be
E_\ell^\pm = \frac 12 \Big( q_0(1+2\delta_\ell)\pm|\vec q\,|\,v\Big).
\en
The zero and first order coefficients $B_0$ and $B_1$ in
Eq.~(\ref{q2Eldistr}) are removed by the folding process since
\be
\int_{E_\ell-}^{E_\ell^+}dE_\ell P_2(\cos\theta(E_\ell))=
\int_{E_\ell-}^{E_\ell^+} E_\ell dE_\ell P_2(\cos\theta(E_\ell))=0
\en
as can be seen by direct calculation or by considering the orthogonality
relations
\be
\int_{E_\ell-}^{E_\ell^+}dE_\ell P_{0,1}(\cos\theta(E_\ell))
  P_2(\cos\theta(E_\ell))=0.
\en
 Similar to Eq.~(\ref{eq:H2}) one finds
\be
\label{Elint}
\frac{d\Gamma_{U-2L}}{dq^2}=   10\int\limits_{E_\ell^-}^{E_\ell^+}\! dE_\ell
  \frac{d^2\Gamma}{dq^2 dE_\ell} P_2(\cos\theta(E_\ell))
  =\frac{2}{2S_1 +1}\frac{\Gamma_0 |\vec q\,|q^2 v^3}{m_1^7}
  {\cal H}_{U-2L}(q^2).
\en
To be sure, we have done the somewhat lengthy $E_\ell$ integration in
Eq.~(\ref{Elint}) and confirmed the expected result on the r.h.s.\ of
Eq.~(\ref{Elint}). From here on one would proceed as in Sec.~\ref{sec:optd},
i.e., one defines an optimized rate by dividing out the
lepton-mass-dependent factor $v^3=(1-m^2_\ell/q^2)^3$. Differing from the
$(q^2,\cos\theta)$ analysis discussed in the main text the $(q^2,E_\ell)$
phase space is not rectangular, which means that the $q^2$ and $E_\ell$
integrations are not interchangeable. The projection of the relevant $B_2$
coefficient Eq.(\ref{Elint}) has to be done for each $q^2$ value, or for
each $q^2$ bin, before $q^2$ integration. In the $\tau$-mode the range of
$E_\ell$ becomes very small near threshold $q^2=m_\tau^2$ and near
the zero-recoil point $q^2=(m_1-m_2)^2$.

In summary, we have proposed a form-factor-independent test of lepton
universality for semileptonic $B$ meson, $B_c$ meson, and $\Lambda_b$ baryon
decays by analyzing the two-fold $(q^2,\cos\theta)$ decay distribution. We
have defined optimized rates for the $e,\mu,\tau$ modes the ratios of which
take the value of 1 in the SM, independently of form-factor effects. The
form-factor independent test involves a reduced phase space for the light
lepton modes which will somewhat reduce the data sample for the light modes.
The requisite angular analysis of the two-fold $(q^2,\cos\theta)$ distribution
will be quite challenging from the experimental point of view. We have
discussed New Physics effects for the $\tau$-mode the inclusion of which
will lead to large aberrations from the SM value of 1 for the ratio of the
optimized rates. As a by-line we have also included a discussion of the 
$(q^2,E_\ell)$ decay distribution as a possible candidate for form factor
independent tests of lepton universality.

We conclude with two remarks.  We have
made a wide survey of polarization observables in semileptonic $b$ hadron
decays to find an observable with the
requisite property that the helicity-flip dependence factors out of the
observable. In fact, in semileptonic polarized $\Lambda_b$ decay one can
identify the observable $v\,({\cal H}_P-2{\cal H}_{L_-})$ which posesses the
desired property~\cite{Kadeer:2005aq,Groote:2019rmj}.
%
%
All in all, we are looking forward to experimental tests of
lepton universality using the optimized branching ratios proposed in this
paper.

\begin{acknowledgments}
J.G.K.\ acknowledges discussions with H.G.~Sander on the experimental aspects
of the problem. M.A.I.\ and J.G.K.\ thank the Heisenberg-Landau Grant for
providing support for their collaboration. The research of S.G.\ was supported
by the European Regional Development Fund under Grant No.~TK133. The research
of S.G.\ and M.A.I. was supported by the PRISMA$+$ Clusters of Excellence
(project No.~2118 and ID~39083149) at the University of Mainz. Both
acknowledge the hospitality of the Institute for Theoretical Physics at the
University of Mainz. The research of V.E.L. was funded  by the BMBF (Germany)
``Verbundprojekt 05P2018 - Ausbau von ALICE am LHC: Jets und partonische
Struktur von Kernen'' (F\"orderkennzeichen: 05P18VTCA1), by ANID PIA/APOYO
AFB180002 (Chile) and by FONDECYT (Chile) under Grant No. 1191103. 
P. S. acknowledges support from Istituto Nazionale di Fisica Nucleare, I.S.
QFT\underline{\hspace*{2mm}}HEP.
\end{acknowledgments}

\end{document}